\documentclass[10pt, conference, compsocconf]{IEEEtran}

\ifCLASSINFOpdf
\else
\fi

\usepackage{cite}
\usepackage{amsmath,amssymb,amsfonts}
\usepackage{graphicx}
\usepackage{textcomp}
\usepackage{xcolor}
\usepackage{gensymb} 
\usepackage{algorithm}
\usepackage{algorithmic}
\usepackage[multiple]{footmisc}
\usepackage{authblk} 
\let\labelindent\relax

\usepackage[shortlabels]{enumitem}

\usepackage{comment} 
\usepackage{float} 
\usepackage{longtable} 
\usepackage{enumitem} 

\def\BibTeX{{\rm B\kern-.05em{\sc i\kern-.025em b}\kern-.08em
     T\kern-.1667em\lower.7ex\hbox{E}\kern-.125emX}}

\begin{document}
%
\title{A Conflict Detection Framework for IoT Services in Multi-resident Smart Homes}

\author[*]{Dipankar Chaki}
\author[*]{Athman Bouguettaya}
\author[$\dag$]{Sajib Mistry}
\affil[*]{School of Computer Science\\ The University of Sydney\\ Australia}
\affil[$\dag$]{School of Electrical Engineering, Computing and Mathematical Sciences\\ Curtin University\\ Australia}
\affil[*]{\textit {\{dipankar.chaki, athman.bouguettaya\}@sydney.edu.au}}
\affil[$\dag$]{\textit {sajib.mistry@curtin.edu.au}}

\renewcommand\Authands{ and }

\maketitle

\IEEEoverridecommandlockouts
\IEEEpubid{\begin{minipage}{\textwidth}\ \\[12pt] \centering
  \copyright 20XX IEEE.  Personal use of this material is permitted.  Permission from IEEE must be obtained for all other uses, in any current or future media, including reprinting/republishing this material for advertising or promotional purposes, creating new collective works, for resale or redistribution to servers or lists, or reuse of any copyrighted component of this work in other works.
\end{minipage}}

\begin{abstract}
We propose a novel framework to detect conflicts among IoT services in a multi-resident smart home. A novel IoT conflict model is proposed considering the functional and non-functional properties of IoT services. We design a conflict ontology that formally represents different types of conflicts. A hybrid conflict detection algorithm is proposed by combining both knowledge-driven and data-driven approaches. Experimental results on real-world datasets show the efficiency of the proposed approach.
\end{abstract}



\section{Introduction}
The Internet of Things (IoT) is fast becoming a pervasive universal computing network where everything and everyone is connected to the Internet \cite{wu2014cognitive}. Technologies such as Wireless Sensor Networks (WSNs), Radio Frequency Identification (RFID) tags form the key framework for the Internet of Things. IoT technology is the key ingredient for cutting-edge applications such as smart campus, smart city, and intelligent transportation system \cite{ranganathan2003infrastructure}.

A pre-eminent application domain for IoT is the \textit{smart home}. A smart home is any regular home that is augmented with IoT devices \cite{huang2018discovering}. The main objective of a smart home is to provide residents with \textit{efficiency} and \textit{convenience} \cite{wu2014cognitive}. The smart home may adjust appliances' settings to suit the residents' habits, thus providing convenience to its residents.

Things in the IoT environment exhibit the same behavior as represented in the \textit{service paradigm}. Each thing has functionality that deliberates with non-functional attributes (QoS). We leverage the service paradigm to abstract the \textit{functional} and \textit{non-functional} properties of smart home devices as \emph{IoT services} \cite{huang2016service}. For example, a light in a smart home is represented as a light service. The functional property of the light service is to provide illumination. Examples of non-functional properties of the light service include luminosity level, power consumption rate, durability.

\IEEEpubidadjcol

There are two types of smart homes: i) \textit{Single resident smart home} and ii) \textit{Multi-resident smart home}. A smart home may recommend services based on a single resident's habits to provide convenience \cite{huang2018convenience}. However, such approaches are not applicable in a multi-resident smart home. In a multi-resident smart home, different residents may have different habits and service requirements, which may lead to \textit{conflicts}. It may not be possible to recommend services without \textit{detecting} and {resolving} conflicts among multiple residents. For example, one resident may prefer the light to be ``on" while watching TV, and another resident may prefer it to be ``off". One resident may prefer the AC temperature to be at 25\degree C, and another resident may prefer it to be at 20\degree C. Hence, a service conflict occurs as the AC or the light cannot satisfy the requirements of multiple residents.


\IEEEpubidadjcol

We focus on the \textit{detection of conflicts} as a prerequisite to resolving conflicts in a multi-resident smart home. A few works focus on designing a comfortable and efficient smart home without considering IoT service conflicts \cite{huang2016service,lin2007multi}. A limited amount of existing literature focuses on conflicts in a context-aware ambient intelligent environment. Systems such as Gaia and CARISMA deal with multi-user conflicts regarding a single application \cite{ranganathan2003infrastructure,capra2003carisma}. A few pieces of research have been conducted to deal with the preference for media applications in a multi-resident smart home \cite{park2005dynamic,shin2009service}. 
The existing frameworks have the following shortcomings:


\begin{itemize}[leftmargin=*]
    \item \textbf{Conflict Ontology}: Ontology is used to categorize things according to their similarities. Ontology is essential to understand the nature of the things and relationships between them \cite{clark1996requirements}. To the best of our knowledge, existing approaches do not focus on designing a conflict ontology.
    \item \textbf{Formal Conflict Model}: Formalism is important for representing knowledge of a specific domain. Expert systems make use of knowledge representation formalisms to discern a particular agent \cite{clark1996requirements}. Existing approaches do not provide formal models for conflict detection.
\end{itemize}

We propose a hybrid approach for conflict detection that combines the strengths of a knowledge-driven and data-driven approach. We consider ontology as a means to represent knowledge that allows us to model different types of conflicts formally. One key aspect of the ontology is the reusability, which is usually preferred in real-world applications \cite{clark1996requirements}. Various IoT application domains such as smart cities, smart campuses can be benefited from the proposed approach.

The key constraint in ontology-based approaches is that it relies on the rigidity of specifications \cite{camacho2014ontology}. If the conflict domain is not appropriately modeled, the smart home system may not be able to detect different types of conflict. For example, the preference for the AC temperature of a resident cannot be modeled beforehand since this type of preference varies from person to person. Service usage history (i.e., previous data) is required to model the preferences. However, the preference model based on a data-driven approach is not reusable, and the conflict model needs to be rebuilt from scratch for a new home. A hybrid approach (using both knowledge-driven and data-driven) is required to construct a conflict detection framework. The key contributions are summarized below:

\begin{itemize}[leftmargin=*]
    \item A conflict ontology that gives explicit and formal definitions of different types of conflicts based on functional and non-functional features of IoT services.
    \item A hybrid conflict detection algorithm that provides the foundation to resolve conflicts.
\end{itemize}

\section{Motivation Scenario}
We discuss two scenarios to illustrate the notion of different types of conflicts in a multi-resident smart home.

\begin{figure}[t!]
\center
\includegraphics[width=.5\textwidth]{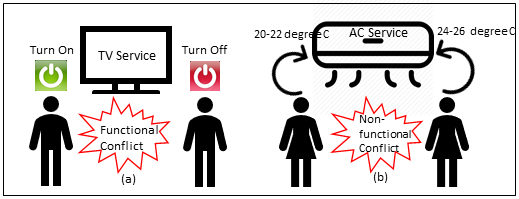}
\caption{Example of conflicting scenarios based on users' requirements.}
\label{motivation}
\end{figure}

\textit{Scenario 1:} Suppose there is only one TV in a home, and a resident, R1, has a habit of studying in the living room, turning off the TV in the evening. Another resident, R2, has a habit of watching TV during the evening. A smart home may understand the residents' habits and may adjust the appliances' settings according to the contexts. Multiple contexts may coexist in an intelligent environment. When R1 and R2 stay together in the living room during the evening, one context would be to keep the TV turned off for R1, and another context would be to turn on the TV for R2. Because of two different contexts, it is not possible to invoke a service by both ``ON" and ``OFF" state at the same time and location. A TV service can serve its functionality when it is ON (i.e., the state of the service is ``ON"). On the contrary, the ``OFF" state denotes that the service is not executing its functionality. A \textit{functional} conflict arises because of the inverse state of the TV service co-occurs (Fig. \ref{motivation} (a)).

\textit{Scenario 2:} Consider a living room equipped with an AC service. A resident, R1, is studying in the living room, and her preferred temperature is between 20\degree C and 22\degree C. Another resident, R2, has a preferred temperature between 25\degree C and 27\degree C in the living room. The smart home may know the contexts of the preferred temperatures of these two residents and may adjust the AC's temperature accordingly. When R1 and R2 stay together in the living room, one context would be to keep the AC temperature between 20\degree C and 22\degree C as per R1's preference, and another context would be to keep the AC temperature between 25\degree C and 27\degree C as per R2's preference. Because of two different contexts, it is not possible to invoke the AC service by two different temperature settings at the same time and location. A \textit{non-functional} conflict occurs since the AC service cannot provide two different temperatures concurrently (i.e., temperature is a non-functional property of an AC) (Fig. \ref{motivation} (b)). For the simplicity of the conflict notion, we assume that the effect of changing the service state is immediate. For example, an AC is executing its operation between 9:00 pm 10:00 pm with 20\degree C temperature. When another request arrives between 10:01 pm and 11:00 pm with 27\degree C temperature, there is no conflict. In practice, the room temperature is not increased by 7 degrees within 1 minute.

A fundamental task is to detect conflicts while recommending services based on different peoples' requirements. For the rest of the paper, we use habits, preferences, and requirements interchangeably. It is possible to understand different peoples' service usage habits by observing their service usage history \cite{chen2015significant}. Usually, a recommender system mines the service usage patterns to automate service provisioning, i.e., making a convenient life. A conflict arises when residents have different requirements for the IoT services, and the service cannot satisfy these requirements at the same time. As a result, there is a need to design a \textit{conflict detection framework} consists of a \textit{conflict model} and a \textit{conflict detection algorithm}.


\section{IoT Service Model}
We represent the notion of \textit{IoT service}, \textit{IoT service event}, and \textit{IoT service event sequences} to illustrate the concept of \textit{IoT service conflict}. We extend the model of an IoT service in \cite{huang2018convenience}. We focus on \textit{shareable} IoT services where conflicts may arise. A \textit{shareable} IoT service serves multiple users at the same time and location. Radio, television, DVD, AC, light, heater, and fan are some examples of shared IoT services. A \textit{non-shareable} IoT service serves only one user at a time. Examples are toaster, microwave oven, electric kettle, and washing machine.

An \textit{IoT Service ($S$)}, is represented as a tuple of \big \langle \textit{$S_{id}$, $S_{name}$, $F$, $Q$}\big \rangle \hspace{0.15 cm}where:

\begin{itemize}[itemsep=0ex, leftmargin=*]
  \item \textit{$S_{id}$} is an unique identifier of the service.
  \item \textit{$S_{name}$} is the name of the service. 
  \item \textit{$F$} is the set of \big \{\textit{$f_1$, $f_2$, $f_3$,.......$f_n$}\big \} where each $f_i$ denotes the function offered by a service. The purpose of having a service is regarded as the functional property of a service.
  \item \textit{$Q$} is the set of \big \{\textit{$q_1$, $q_2$, $q_3$,.......$q_m$}\big \} where each $q_j$ denotes a QoS attribute of a service.
\end{itemize} 

For example, a TV service can be represented as \big\langle\textit{5, TV, \{telecasting programs, receptor for security camera\}, \{$\infty$, \$600, 200 Watts, 5 years\}}\big\rangle. Here, 5 is the id of the TV service by which it can be uniquely identifiable. The name of the service is TV. Functional properties are telecasting programs such as news, sports, and it could act as a receptor for a security camera. \{$\infty$, \$600, 200 Watts, 5 years\} represents the non-functional properties such as capacity, price, electricity consumption rate, and warranty of the service, respectively. Here, the capacity of the TV service is denoted as $\infty$, meaning the resource capacity is undefinable. It can serve multiple people at the same time and location by telecasting programs.

An \textit{IoT Service Event ($SE$)}, is an instantiation of a service. When a service manifestation occurs (i.e., turn on, turn off, increase, decrease, open, close), an event records the state of the service along with its functional and non-functional properties, user of the service, service execution time and location of the service. A service event can be represented as a tuple of \big \langle \textit{\{$S_{id}, F, Q\}, T, L, U$}\big \rangle \hspace{0.15 cm}where:

\begin{itemize}[itemsep=0ex, leftmargin=*]
  \item \textit{$S_{id}$} is the unique identifier of the service that has some functional ($F$) and non-functional ($Q$) properties.
  \item \textit{$T$} is the execution time of the service. It is represented as a set  \{\textit{$T_s$,$T_e$}\} where $T_s$ denotes the service start time and $T_e$ denotes the service end time.
  \item \textit{$L$} is the location of the service.
  \item \textit{$U$} is the user of the service.
\end{itemize} 

An example of the IoT Service event is \big\langle \textit{\{5, \{telecasting programs\}, \{sports, 35 dB\}\}, \{07:45, 08:45\}, living room, 3}\big\rangle. Here, 5 is the id of the TV service by which it can be uniquely identifiable. 07:45 is the service start time, and 08:45 is the service end time. The execution time of the service is 08:45-07:45=1 hour. During the execution time, it's functionality is telecasting programs. \{sports, 35 dB\, $\infty$\} represents the non-functional properties such as channel (sports), volume (35 dB) and capacity ($\infty$) of the service. Living room denotes the location where the service operates its functionality. 3 is the unique identifier of the resident who is using the service.

Usually, residents interact with IoT services for various household chores. These interactions are recorded as \textit{IoT Service Event Sequences ($SES$)}. Service Event Sequences is a set of \big \{\textit{$SE_1$, $SE_2$, $SE_3$,.......$SE_k$}\big\} where each $SE_i$ is an IoT service event.

An IoT service, $S$, is typically associated with a set of functional and non-functional properties. A service can be used distinctively by different residents. Different residents' service usage requirements can be captured from IoT service events, $SE$. An IoT service event demonstrates how a resident is using a particular service, along with time and location. The history of service events is stored in a database called IoT service event sequences, $SES$. Different residents may have different requirements to use a service that may cause a conflict, $Conf$. The paper aims to identify a function $F(S, SES)$, where $Conf \approx F(S, SES)$. In other words, our goal is to detect conflict using service-related and usage-related data.

\section{IoT Service Conflict Detection Framework}

\begin{figure}[htbp]
\center
\includegraphics[width=\linewidth]{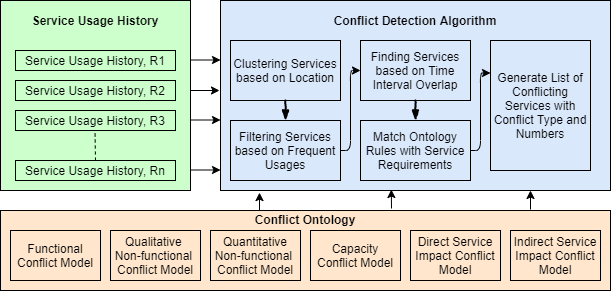}
\caption{IoT service conflict detection framework.}
\label{framework}
\end{figure}

The proposed IoT service conflict detection framework has three major components: \textit{service usage history}, \textit{conflict ontology}, and \textit{conflict detection algorithm} (Fig. \ref{framework}). Service usage data represent the requirements of the residents. Conflict ontology is the foundation of conflict detection that defines various types of conflicts. The conflict detection algorithm matches ontology with the usage data and captures the type of conflicts.

\subsection{Service Usage History}
Service usage data records the residents' interaction with the services. It is possible to understand different peoples’ service usage preferences by observing their service usage history. Different residents may have different preferences over a service usage, which may cause a conflict.

\subsection{IoT Service Conflict Model}
Conflict is a natural disagreement between different attitudes, beliefs, values, or needs \cite{wang2011development}. Tuttlies et al. define conflict concerning a user or application as \textit{``...a context change that leads to a state of the environment which is considered inadmissible by the application or user"} \cite{tuttlies2007comity}. However, the definition of conflict varies from application to application based on the context. We are interested in IoT service conflict when different people have different requirements to use a particular service. An \textit{IoT Service Conflict} occurs when an individual service or multiple services cannot satisfy the requirements of multiple users at the same time duration and location.

\begin{figure}[htbp]
\center
\includegraphics[width=.49\textwidth,height=8cm,keepaspectratio]{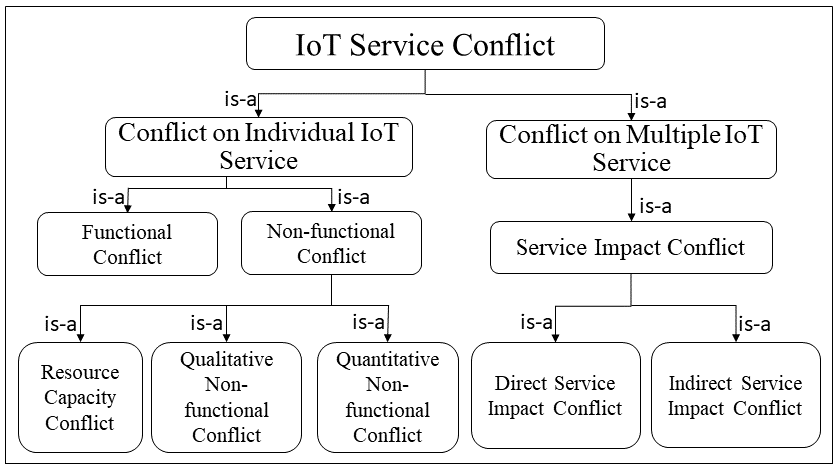}
\caption{IoT service conflict ontology.}
\label{ontology}
\end{figure}

We propose an IoT service conflict ontology (Fig. \ref{ontology}). The primary parent of the conflict ontology is divided into two categories, whether the conflict occurs on an \textit{individual service} or the conflict occurs among \textit{multiple services}. It is of paramount importance to figure out a few criteria without which conflict cannot be modeled appropriately. Conflicts are defined based on the service usage requirements of each resident, and these requirements can be gathered from the IoT service event. Given two service events ($se_i$,$se_j$), the following conditions have to be satisfied to be considered as a conflict situation.
\begin{itemize}[itemsep=0ex, leftmargin=*]
    \item $l_{s_i}$ $\simeq$ $l_{s_j}$, denoting that the two services ($s_i$, $s_j$) are executed in the same location.
    \item $(st_i, et_i) \cap (st_j, et_j)) \neq \emptyset$, meaning that two service events ($se_i$,$se_j$) are invoked simultaneously and there is a temporal overlap between them. We use Allen's temporal relation to detect overlapping events \cite{allen1994actions}.
    \item $u_{s_i} \neq u_{s_j}$, denoting that these two events are invoked by different users.
\end{itemize}

After fulfilling these criteria, each type of conflict is formalized below. Since the ontology has a hierarchical tree structure where each child class is a sub-class of its parent class, only the bottom classes of each branch are formalized. A parent class can be formalized by the conjunction operation of the child classes.

\subsubsection{Functional Conflict}

A functional conflict arises when different residents have different state requirements on the functional property of a service concurrently. The requirements of the service state must be the opposite, e.g., turn-on and turn-off. Given two service events $se_i$, $se_j$ where $s_i$ and $s_j$ denote the same service with the services' functional state requirements ($s_i.st$,$s_j.st$), functional conflict ($fc$) is:

\small{
\begin{equation}
    fc \implies (s_i.st(on) \land s_j.st(off)) \lor (s_i.st(off) \land s_j.st(on))
\end{equation}
}



\subsubsection{Non-functional Conflict}

A non-functional conflict arises when residents have different Quality of Service (QoS) preferences for the same IoT service, and these preferences cannot be satisfied at the same time. The non-functional conflict is further divided into \emph{resource capacity conflict}, \emph{qualitative non-functional conflict}, and \emph {quantitative non-functional conflict}.

\paragraph{Resource Capacity Conflict}
A resource capacity conflict occurs as a result of not having sufficient capacity to use an IoT service by more than one user. Given two service events $se_i$, $se_j$ where $s_i$ and $s_j$ denote the same service with the services' required capacity ($c_{si}$,$c_{sj}$), resource capacity conflict ($rcc$) is denoted as:

\small
\begin{equation}
    rcc \implies (c_{si} + c_{sj}) > s.c
\end{equation}



\paragraph{Qualitative Non-functional Conflict}
A qualitative non-functional conflict occurs when different people have different nominal QoS preferences over a single service at the same time duration and the same location. Given two service events $se_i$, $se_j$ where $s_i$ and $s_j$ denote the same service with the services' qualitative non-functional attributes ($s_i.qlnf$,$s_j.qlnf$), if there exist at least one property which is different between $s_i.qlnf$ and $s_j.qlnf$, a qualitative non-functional conflict is arised. A qualitative non-functional conflict ($qlnfc$) is defined as:

\small
\begin{equation}
    qlnfc \implies \exists qlnf_k \in s.qlnf: s_i.qlnf_k \neq s_j.qlnf_k
\end{equation}


\textit{Quantitative Non-functional Conflict} A quantitative conflict occurs when different people have different numeric QoS preferences over a single service at the same time and the same location. Given two service events $se_i$, $se_j$ where $s_i$ and $s_j$ denote the same service with the services' quantitative non-functional properties ($s_i.qnnf$,$s_j.qnnf$), quantitative non-functional conflict ($qnnfc$) is denoted as:

\small
\begin{equation}
    qnnfc \implies \exists qnnf_k \in s.qnnf: s_i.qnnf_k \neq s_j.qnnf_k
\end{equation}


\subsubsection{Service Impact Conflict}
A service impact conflict arises when multiple residents prefer to access multiple IoT services, and these services have an impact on each other directly (through the functional property) or indirectly (through the non-functional property). The service impact conflict is further divided into \emph{direct service impact conflict}, and \emph {indirect service impact conflict} as follows.

\paragraph{Direct Service Impact Conflict}
A direct service impact conflict occurs when multiple residents like to access multiple services that have an impact on each other's functional property. For that, these services cannot function properly to satisfy the requirements of the residents. Given two service events $se_i$, $se_j$ where $s_i$ and $s_j$ denote different services (but, they have functional dependency on each other and we use $\subset$ symbol to represent dependency) and their functional states ($s_i.st$,$s_j.st$), a direct service impact conflict ($dsic$) is:

\small
\begin{equation}
    dsic \implies ((s_i \subset s_j) \lor (s_j \subset s_i)) \land (s_i.st(on) \land s_j.st(on))
\end{equation}


\paragraph{Indirect Service Impact Conflict}
An indirect service impact conflict occurs when multiple residents prefer to use multiple services that have an impact on each other's non-functional property. For that, these services cannot satisfy the needs of the residents. Given two service events $se_i$, $se_j$ where $s_i$ and $s_j$ denote different services (they do not have any functional dependency on each other) and their non-functional properties ($s_i.nfp$,$s_j.nfp$), an indirect service impact conflict ($isic$) can be defined as:

\small
\begin{equation}
    isic \implies ((s_i \not\subset s_j) \lor (s_j \not\subset s_i)) \land (s_i.nfp \simeq s_j.nfp)
\end{equation}


\subsection{Conflict Detection Algorithm}

We develop an algorithm to detect conflicts from the IoT service event sequence dataset. The algorithm is based on frequent service usage and time interval overlap. The details of the proposed approach have mainly two phases: Phase 1 and Phase 2 are illustrated in Algorithm 1 and Algorithm 2, respectively.

\subsection*{Phase 1: Service Selection and Set of Intervals Creation}
The algorithm aims to select top-k services based on their usage frequency. The intuitive idea of our heuristic is, the services which are used more frequently by the residents, are more prone to have conflicts. Then, we create a set of intervals for the selected services. The input of this algorithm is the dataset of the service event sequences $SES$ and $k$. Each event sequence ($SE$) contains service ID ($S$), service name ($SN$), location of the service ($L$), service start time ($T_s$), service end time ($T_e$), service user ($U$) along with the service's functional ($F$) and non-functional attributes ($Q$). The system expert gives the value of $k$.

\begin{algorithm}[t!]
\small
\caption{Service Selection and Set of Intervals Creation}\label{alg:algorithm1}
\begin{algorithmic}[1]
\REQUIRE
$SES, k$
\ENSURE
$OV$

\STATE {$LS = \emptyset, CS = \emptyset, TM = \emptyset, SV = \emptyset, TI = \emptyset, OV = \emptyset$}

\item[] // Clustering services located in a same location
\FOR{\textbf{each} $se_i$ in $SES.SE$}
\FOR{\textbf{each} $l_i$ in $SE.L$}
\FOR{\textbf{each} $s_j$ in $SE.S$}
\IF{$s_j.l$ is equal to $l_i$}
\STATE $LS_i \leftarrow insert(s_j)$
\ENDIF
\ENDFOR
\ENDFOR
\ENDFOR
\item[] // Selecting frequent services based on usage
\FOR{\textbf{each} $l_i$ in $LS$}
\FOR{\textbf{each} $s_j$ in $l_i$}
\STATE $CS \leftarrow count(s_j)$ 
\STATE $TM \leftarrow addTimeInterval(s_j.T_s, s_j.T_e)$
\ENDFOR
\ENDFOR
\STATE $SV \leftarrow sort(CS)+sort(TM)$
\STATE $SS \leftarrow top (SV,k)$
\STATE $TI \leftarrow timeInterval(SS)$
\STATE $OV \leftarrow overlap(SS)$
\RETURN OV

\end{algorithmic}
\end{algorithm}

A smart home usually has several locations, such as a living room, a kitchen, a bathroom, and a bedroom. Each IoT service is associated with a location. For instance, a TV is located in a living room; a toaster service is located in a kitchen. We cluster services that are located in the same location (lines [2-10] in algorithm 1). When a TV service and a DVD service are located in a living room, service cluster of the living room is $LS_{living} = \big \langle S_{TV}, S_{DVD} \big \rangle$.

For each service located in a location, we count the number of times ($CS$) and amount of times ($TM$) they have been used (lines [11-16] in algorithm 1). We sort their values and combine them to get the set ($SV$) of services along with their frequency (line 17). Then, we select top-k used services ($SS$) (line 18). In Australia, a home has 17 connected devices (i.e., services) on average and 2.6 residents\footnote{https://tinyurl.com/vg3lh6w}\footnote{https://tinyurl.com/y6jzjwf4}. We assume that each resident uses $17/2.6=6.5$ devices. For that, we initialize $k$ as 7. When a house has $n$ services that is less than 7, we consider $k = min(n,k)$.

We know that each service has an execution time. We create a set of intervals ($TI$) for each selected service (line 17). For example, on June 15, 2018, one resident turned on a light service, $S_{light}$ at 1 pm and turned it off at 3 pm. He also turned on the same service at 6 pm and turned it off at 8 pm. On the same day, another resident turned on the same service at 2 pm and turned it off at 4 pm. He also turned on the same service at 4 pm and turned it off at 7 pm. A set of time intervals for light service will be $S_{light} =  \{ \langle 1, 3 \rangle, \langle 6, 8 \rangle, \langle 2, 4 \rangle, \langle 4, 7 \rangle  \} $. We find overlapping services based on time intervals (line 18). A service having requests with overlapping time intervals is regarded as overlapping services. On each index of $OV$, there are two services.

\subsection*{Phase 2: Detection of Service Conflicts}

The input of this algorithm is the set of overlapping services ($OV$) from Phase 1 and the set of rules ($R$) from the formal conflict model (Section IV-A). If the ontological rule matches the overlapping services, then a conflict is found, and the list of conflicting services is updated. The outcomes are the number of conflicts ($CN$), the type of conflicts ($CT$), and a set of conflicting services ($CS$).

There are some cases when ontological rules cannot detect conflict properly. For example, one resident's preference for AC temperature is 22\degree C, and another resident's preference is 23\degree C at the same time and location. This scenario is counted as a conflict according to our ontology since 21\degree C is not equal to 22\degree C. In practice, 1\degree C doesn't make any major difference. The key technique of the hybrid approach is to find a range for numerical data according to each resident's preference. When a service receives a request that falls outside of that range, we define it as a conflict. For example, a resident is sleeping, keeping the temperature at 18\degree C. If another resident comes to the same place whose preference range is [20-22\degree C], there is a conflict. We calculate the range using standard deviation ($\sigma$) and median ($m$). The standard deviation measures the spread out of the numbers (e.g., preferred temperature), and the median measures the middle value of the resident's preferred temperature from the history. We calculate the range, $Rn = (m-\sigma, m+\sigma$) where $m-\sigma$ is the minimum value and $m+\sigma$ is the maximum value. The hybrid approach is useful when there is a possibility to have confict between quantitative (i.e., numerical) attributes of the dataset. When the attribute of the dataset is nominal, the proposed ontology model works properly.

\begin{algorithm}[t!]
\small
\caption{Detection of Service Conflicts}\label{alg:algorithm2}
\begin{algorithmic}[1]
\REQUIRE
$OV, R\{R_{fc}, R_{qlnfc}, R_{qnnfc}, R_{rcc}, R_{dsic}, R_{isic}\}, R_n$
\ENSURE
$CN$,$CT$,$CS$

\STATE ${CN = 0, CT = \emptyset, CS = \emptyset}$

\FOR{\textbf{each $r_j$ in $R$}}
\FOR{\textbf{each} $ov_i$ in $OV$} 
\IF{$ov_i$ matches $r_j$}
    \item[] // Checking whether the rule is for the quantitative conflict
    \IF{$r_j$ is equal to $r_qnnfc$}
    \STATE $r_j.Conflict \leftarrow R_n$ //apply preference range to the conflict checking

    \ELSE
        \STATE $CT \leftarrow r_j.Conflict$
        \STATE $CN_{CT} += 1$ 
        \STATE $CS_{CT} \leftarrow insert(ov_i)$
    \ENDIF
\ENDIF
\ENDFOR
\ENDFOR
\STATE return $CN$,$CT$,$CS$
\end{algorithmic}
\end{algorithm}



After detecting a conflict, we calculate conflict weight ($w$) for that particular type of conflict. The conflict weight represents the time proportion of a conflict. For example, one resident wants to watch the news channel in a TV service between 07:30 pm and 08:30 pm ($ti_1 = 60$). Another resident wants to watch the sports channel between 08:10 pm and 08:40 pm ($ti_2 = 30$). Between $ti_1$ and $ti_2$, maximum time duration  is $60$ and overlapping time is $20$. According to Eqn. 7, $w = 20/60$ becomes $0.333$.

\begin{equation}
    w = (ti_1 \cap ti_2)/maxDuration(ti_1,ti_2)
\end{equation}

\section{Experimental Results and Discussion}

\subsection{Experiment Setup}
The proposed approach has been evaluated using a real dataset collected from the Center for Advanced Studies in Adaptive Systems (CASAS) \cite{cook2012casas}. We use java programming language, and the experiment is performed on a 3.20GHz Intel(R) Core(TM) i7-8700 CPU and 16 GB RAM under Windows 10 64-bit Operating System. There are a few multi-resident activity datasets available. However, those datasets are not useful as they do not have any conflicting situations of IoT service usage. Multi-resident activity datasets reflect compromises of tenants interacting with services. Thus not showing conflicts. In contrast, records of single-resident interactions with IoT services show the actual preferences of an individual for what conflicts may arise. With that respect, we use the service interaction records of four individual residents. The dataset contains different types of sensors. They are battery level sensors, magnetic door sensors, light switches, light sensors, infrared motion sensors, and temperature sensors. In this study, we consider each \textit{sensor} as an \textit{IoT service}. Table \ref{tab1} shows the description of all the attributes in our dataset.

\begin{table}[htbp]
\caption{Description of the attributes of the dataset.}
\label{tab1}
\begin{tabular}{|l|l|p{6.7cm}}
\hline
\textbf{Attributes} &  \textbf{Description} \\
\hline
Date &  The service execution date\\ \hline
Time &  The service execution time\\ \hline
Sensor & \parbox{6.7cm}{Name of the sensors such as door sensors, light switch sensors, motion sensors, light sensors, temperature sensors}\\\hline
Status & \parbox{6.7cm}{ON, when the service starts executing, and OFF, when the service stops executing}\\
\hline
\end{tabular}

\end{table}

The dataset contains two months of usage data between June 15, 2011, and August 14, 2011. We use \textit{sensor} and \textit{IoT service} interchangeably for the rest of the paper. The dataset only contains the on/off time of the sensors. We consider ``ON" as a service start time and ``OFF" as a service termination time. We only could detect functional conflicts from this dataset. Other attributes, such as QoS properties are absent in this dataset. A dataset is augmented to test our approach based on some QoS attributes of IoT services.

\subsection{Performance Evaluation}

We first examine the effectiveness of the proposed hybrid approach. The augmented dataset contains 1000 rows of service events (mainly, it contains usage history of TV, AC, light, and window). The dataset is split into two sets: (1) 80\% is used for training the model, and 20\% is used to compute their accuracy. Precision, recall, f1-score, and accuracy are computed to evaluate the performance of the framework. The dataset consists of two months of thermal preference records of two residents. Among them, there are 352 overlapping service events of AC. The ground truth models the thermal preference, as described in \cite{humphreys2002validity}. According to the study, approximately 3\degree C is the difference between individual residents' comfortableness. For example, when one resident has a preference of 22\degree C, and another resident has a preference of 25\degree C, they are comfortable. If the difference is more than 3\degree C, they are not comfortable. We assume, if they are not comfortable, there is a conflict.

\begin{figure}[htbp]
\begin{center}
\includegraphics[width=0.85\columnwidth]{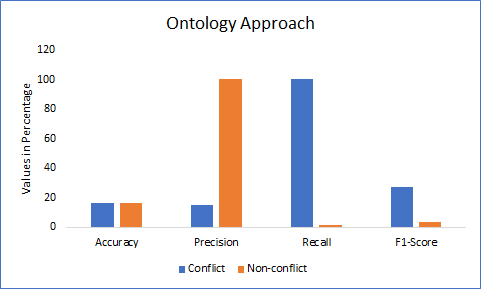}
\caption{Accuracy, precision, recall, f1-score in ontology approach.}
\label{fig3}
\end{center}
\end{figure}

\begin{figure}[htbp]
\begin{center}
\includegraphics[width=0.85\columnwidth]{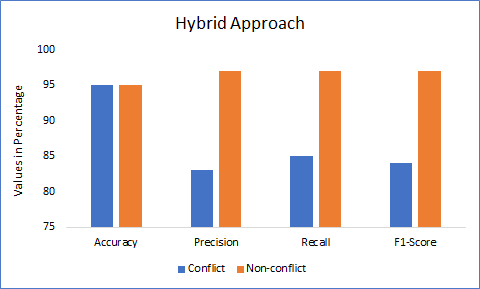}
\caption{Accuracy, precision, recall, f1-score in hybrid approach.}
\label{fig4}
\end{center}
\end{figure}

It is visible from fig. \ref{fig3} and fig. \ref{fig4} that the hybrid approach can detect both conflict situations and non-conflict situations more accurately than the ontology-based approach. According to the ground truth, there are 53 conflict instances and 299 non-conflict instances among the 352 overlapping events. To evaluate the proposed model, we only consider non-functional quantitative conflict (i.e., temperature preference). The proposed hybrid approach accurately detects 45 instances as conflicts and 290 instances as non-conflicts (accuracy 95\%). On the contrary, the ontology approach detects 53 instances as conflicts. However, it detects 294 instances as conflicts, whereas these instances are non-conflicts. When the temperature difference is even 1\degree C, it detects as a conflict according to the ontology. However, the ground truth depicts if the difference is more than 3\degree C, then there is a conflict. Our hybrid approach detects both conflicts and non-conflicts more accurately based on precision, recall, and f1-score.

\subsection{Complexity Analysis}
The proposed heuristic scheme has a similar polynomial computation complexity like brute-force when there are a few numbers of services. Our approach compares the ontological rules with service usage data. If there are $n$ services and $m$ number of events, the time required to check conflict is $\mathcal{O}(n*m)$ for both approaches. However, our heuristic approach works efficiently when there are many services in a home, but, only a few services (top-k) are used frequently.

\begin{figure}[htbp]
\begin{center}
\includegraphics[width=0.85\columnwidth]{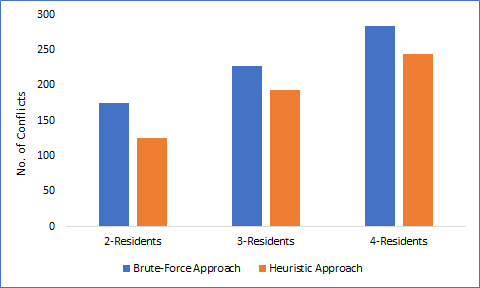}
\caption{No. of conflicts between residents.}
\label{fig5}
\end{center}
\end{figure}

\subsection{Comparison}
We conduct another experiment to test our proposed heuristic approach based on the hybrid-approach. We detect the number of functional conflicts since the real dataset only contains information on the sensor start time and end time. Figure \ref{fig5} demonstrates that our heuristic approach has less accuracy than the brute-force approach. The details of the heuristic approach are described in algorithm 1 (line 18). We conduct this experiment considering conflicts between 2-residents, 3-residents, and 4-residents, respectively. The performance of heuristic is better in terms of time, as it searches only the selected services (in algorithm 2) that are more frequently used. However, we focus more on accuracy in this paper. We remove k from line 18 in algorithm 1 and select all the services to compare our brute-force approach with other rule-conflict detection algorithms such as UTEA \cite{sun2014conflict} and IRIS \cite{shehata2007using}. For comparing several types of conflict, we use the augmented dataset introduced earlier. The dataset contains non-functional attributes such as light luminosity, TV volume, TV channels, AC temperature, window opening/closing time. The comparison results are demonstrated in table \ref{tab2}.

\begin{table}[htbp]
\caption{Comparison with UTEA and IRIS}
\label{tab2}
\begin{tabular}{|l|l|l|l|} 
\hline
\textbf{Conflict Type} & \textbf{Our Work} & \textbf{UTEA} & \textbf{IRIS}\\
\hline
Functional & 778 & 778 & 778\\ \hline
Capacity & 32 & 0 & 0\\ \hline
Qualitative Non-Functional & 267 & 267 & 267\\ \hline
Quantitative Non-Functional & 45 & 72 & 0\\ \hline
\end{tabular}
\end{table}

IRIS mainly detects conflict that are known as device conflict with opposite functionality. This is why, it detects accurate number of functional and qualitative conflicts. However, it does not consider conflict on environmental variable such as temperature, light, etc. On the contrary, UTEA can capture more complex conflicts such as quantitative and direct impact conflict. However, its miss-detection rate for capturing quantitative conflict is high as it does not use previous data to model preferences (i.e., temperature preference). It captures 72 quantitative non-functional conflicts where the ground truth has 53 conflicts. 



\section{Related Work}
Relevant literature regarding different notions of conflict and their detection techniques are surveyed, which include their sources and solvability. Conflicts can occur based on the following categories: (1) time of detection, (2) source, (3) intervenient, and (4) solvability.

The time when a conflict is detected is a crucial factor in a home automation system. A conflict can be anticipated early (i.e., prediction of a conflict). This is called an apriori (i.e., potential) conflict \cite{syukur2005methods}. The potential conflict has two types: (1) a definite potential conflict, and (2) a possible potential conflict. The former defines a conflict that will occur if the user is in the right context. The latter defines a conflict that may still not occur, even though the user is in the right location and time.

A conflict may occur based on four types of sources \cite{xu2009ontology}. When multiple users concur over a given resource, e.g., a TV, a conflict may occur, which is known as a resource level conflict. When multiple applications concur over a resource, e.g., building management applications trying to control a room’s lighting, a conflict may occur and is known as an application-level conflict. When conflict is raised due to the conflict policies for a given context, a conflict may occur and is known as a policy level conflict. For instance, a user enters a library with a smartphone. He can listen to music through the smartphone's speakers, but the library has a silence policy that requires him not to create noise via speakers. A conflict may occur at the profile level. When there are different user preferences at the same context, e.g., one user prefers to read with the lights to be at full capacity, and another user prefers to watch TV with the light to be at half capacity.

Conflicts may happen because of intervenients \cite{yagita2015application}. Even in a single-resident home, a  conflict may occur based on conflicting intentions, e.g., saving energy and comfort. When there are two or more residents live in a home, a conflict may arise because of concurrency over a resource. A conflict may occur between resident and space, where a user’s actions conflict with any established space policies, e.g., a user’s smartphone ringing in a room with a silence policy.

Finally, conflict detection can be categorized by their solvability \cite{goynugur2016automatic}. Conflict prevention is the best possible solution. When detection happens before its actual occurrence, a conflict can be solved beforehand, which is known as conflict prevention. In general, a conflict is detected during its run time (actual occurrence), and a smart home tries to resolve that conflict. When the system fails to resolve it automatically, it may acknowledge that it is unable to deal with the conflict.

A service-centric feature interaction framework has been presented for integrated services in \cite{igaki2010modeling}. The authors model each service or appliance as an object consists of methods and properties. The states of the services have been characterized as properties, and functionality has been abstracted as methods. A conflict may arise on a service object or an environment object when multiple methods try to update the same properties of the object \cite{metzger2003feature}. The major contribution of this study is to provide guidelines in such a way that undesirable conflicts can be avoided.

Another probable situation is, when the system cannot detect the conflict at the early possible time and later, realizes that it is happened due to delayed sensor information \cite{al2019iotc}. The system must have the ability to acknowledge its occurrence by informing the users of the occurred conflict. As mentioned above, conflicts can be of various types and arise for many reasons. Therefore, different conflict detection and resolution techniques are required.


\section{Conclusion and Future Work}
We propose a hybrid framework combining both knowledge-driven and data-driven approaches to detect conflict among IoT services in multi-resident smart homes. The framework provides a foundation for conflict resolution and to recommend convenient and efficient IoT services. The experimental results show the effectiveness and efficiency of the proposed approach. Our future work focuses on the resolution of different types of conflicts and building a recommender system for multi-resident smart homes.

\def\IEEEbibitemsep{7pt plus 0.5pt}
\bibliographystyle{IEEEtran}
\bibliography{ConfDet}

\end{document}